\documentstyle[prd,aps,preprint,epsfig]{revtex}
\begin{document}
\preprint{\parbox{5cm}{MPG--VT--UR 261/05}}
\title{
Brightness constraint for cooling models of\\ young neutron stars}
\author{Hovik  Grigorian \protect\cite{address1}}
\address{Institut f\"ur Physik, Universit\"at Rostock, D-18051 Rostock,
Germany}
\maketitle
\begin{abstract}
We study the systematics of neutron star cooling curves with three
representative masses from the most populated interval of the
estimated mass distribution for compact objects. The cooling
simulations are made in the framework of the {\it nuclear medium
cooling} (NMC) scenario using different combinations of possible
nucleon-nucleon pairing gaps. Possible heating or enhanced cooling
mechanisms in the crust are not considered. We define a constraint
on the highest possible temperatures for a given age of young
neutron stars and show that this limits the freedom of modeling
pairing gaps and crust properties.


\end{abstract}

Studies of neutron star cooling evolution become very actual due
to the presently known surface temperature and age data provided
by X-ray observatories such as CHANDRA, XMM Newton and from the
ROSAT catalogue \cite{rosat}. These new data open a wide
perspective for nuclear astrophysics for which up to now the
knowledge of internal structure of the compact stars and the
properties of stellar matter under extreme conditions remain
central problems. Theoretical models and hypotheses about the
equation of state of high density matter provide different
alternatives for the explanation of the same set of observational
{\it temperature - age} (TA) data points, when additional
constraints are not provided. In this work we point out an
existing correlation between the crust model and cooling behaviour
of light neutron stars, which has a selective power in combination
of
TA data with the mass spectrum of neutron stars.

In our recent investigations of the cooling evolution of neutron
stars (NS) we have adopted the so called {\it nuclear medium
cooling} (NMC) scenario \cite{Blaschke:2004vq}, which goes beyond
the {\it minimal cooling scenario} \cite{Page:2004fy}, where
in-medium modifications of cooling regulators by definition have
been disregarded. Both approaches agree in the philosophy that
such very effective cooling mechanism like the direct Urca process
should not occur in typical NS. In these approaches the main
cooling process is the modified Urca process, which in our NMC
scenario also includes the in-medium softening of the pion
propagator \cite{Voskresensky:2001fd}. Earlier investigations
within this cooling scenario\cite{SVSWW97,Blaschke:2000dy} have
chosen the crust model as a simplified  {\it Tsuruta law} $ T_{\rm
s}^{\rm Tsur}=(10 ~T_{\rm in})^{2/3}$. Although it is shown in
Ref. \cite{Schaab:1999as} that the cooling evolution could be
essentially affected by the inclusion of internal heating,
nevertheless the latter is expected to be important for late time
evolution and will not affect the results of this paper. It has
been omitted in the present cooling scenario. For more recent
reviews on the cooling scenarios see
\cite{Yakovlev:2004iq,Page:2005fq,Sedrakian:2006mq}.

The cooling simulations presented in this work are based on a code
with a number of improved inputs concerning the heat conductivity,
the nucleon-nucleon pairing gaps and a new model of the neutron
star crust and envelope. These models are basically taken from the
recent calculations of Ref. \cite{Yakovlev:2003qy}, where the
amount of light elements in the crust and the influence of the
magnetic field have been taken into account.

The choice of the structure of crust and envelope of the compact
object becomes a central question due to the direct connection of
the surface temperature with the observations.

Since simultaneous measurements of the surface temperature and either
the mass or the radius of neutron stars are absent, one needs to
relate the known TA data  with other observational
information in order to discriminate between different models.

One quantitative approach has been suggested in \cite{p03,p04} and
developed recently for the NMC scenario \cite{Popov:2004ey}. It
uses the statistical method of population syntheses in combination
with models for the cooling evolution and the mass distribution of
isolated objects to derive a Log N - Log S dependence to be
compared with observational data. This allows for an additional
selection among those scenarios which have successfully passed the
TA test. Nevertheless, in the routine of the Log N - Log S test
the early stages of the time evolution are not included, but only
the tracks of relatively old objects ($t \gtrsim  3 \cdot 10^{4}$
yr). This period is already in the transition from the neutrino to
the photon cooling era. We will show in this work that the cooling
behavior of the young objects is strongly affected by crust
properties, which however do not alter the Log N -Log S
distribution.

In the spirit of the Log N - Log S test we suggest an additional
condition to the TA test, which requires young neutron stars ($
10^3 \leq $ t/yr $\leq 3 \cdot 10^4$) with low masses $\sim 1.1
M_{\odot}$ not to have temperatures (i.e. brightness = luminosity
/ 4 $\pi$ distance$^2$, where absorption could be neglected for
hot sources since they have a hard spectrum) exceeding those of
presently known objects of the same age. We will therefore call it
a {\it brightness constraint} (BC).

 Therefore, it is unlikely that objects with a given age are
hotter than those already observed. As a consequence, cooling
models which would predict such objects should be rejected.

Our approach is based on the following assumptions
   made on the basis of current observational data:
\begin{itemize}
  \item {\it All objects with temperatures exceeding T$=4 \cdot 10^6$ K
are potentially observable.}

 The probability to find an object
increases with its luminosity due to the increase of the
visibility volume and reduction of the absorption of X-rays by the
interstellar medium (see \cite{Wilms:2000ez}). The absorption
falls by several orders of magnitude with increase of the photon
energy from 0.1 to $\sim 10$ keV.
   One can estimate the lower limit of the temperature of an object at the
distance of 10 kpc or even more distant and behind clouds of the
interstellar medium with maximal absorption, say for hydrogen column
densities $n_H \sim 2 \cdot 10^{22}$ cm$^{-2}$ (corresponding to
galactic coordinates l=30, b=0), which would be observed with
instruments like ROSAT, with a lower limit for the brightness given by
a few $0.01$ cts/s. Using such an estimate NASA's HEASARC tools web-page
\footnote{http://heasarc.gsfc.nasa.gov/docs/tools.html}, one obtains a
limiting temperature $T=4 \cdot 10^6$ K, close to the temperature of
Crab pulsar ($2.2 \cdot 10^6$ K).
Therefore, in a good approximation the temperature
of the observed object with highest temperature could be
considered as a border of BC.

   To specify the BC as a test we have made further assumptions:

  \item {\it The distribution of young objects is a step-like function of
  temperature for higher temperatures.}

  \item {\it The uncertainty of the BC border is of the same order
as the error bars of the measured temperatures at the corresponding age.}
\end{itemize}

  We assume that with sufficient observational data it is possible
to measure the temperature distribution of the young objects.
  Even if the population of young objects is small ($\bar N \simeq$
100 \cite{Popov:2006ma}) one can not conclude that the probability
to find an object with arbitrary high temperature is finite. The
distribution is likely a step like or very stiffly falling
function of temperature for higher temperatures, because the star
can not have arbitrarily high temperature from the beginning of
the evolution and the mass distribution of the compact objects are
also a step like function for the small masses (see
\cite{Popov:2004ey}), which are expected to populate the domain of
slower coolers.

 The statistical analysis to define the BC border is complicated by
the small total number of potentially observable objects. It is
assumed to be $\simeq 0.1* \bar N$ \cite{Faucher-Giguere:2005ny},
because all known objects are colder than $4 \cdot 10^6$ K
\cite{Kaplan:2004tm}.

Therefore we leave a corridor between the upper limits of the
observed data points of higher temperatures at  a given age and the
BC border on the 2$\sigma$ level of the error bar of measured
temperature where it was possible and put $\delta~log_{10}(T/K)
\simeq$ (1/$\bar N) *$ ( the whole expected temperature interval
of young objects)/(the probability to be observed) $\simeq
0.4*0.01/0.1 = 0.04$ (for Crab), which is of the same order as the
error bars of measured temperatures.

This last estimation is based on the assumption that the probability
for the existence of an object is equally distributed for all available
temperatures.

In order to apply this simple BC test to already published
\cite{Blaschke:2004vq} scenarios we choose the
strategy to follow the cooling evolution of objects
with the representative values $ M = 1.1, 1.21,$ and $1.41 M_\odot$
corresponding to the most populated bins in the mass
spectrum of Fig. 1 in \cite{Popov:2004ey}.
The basic idea behind this mass spectrum of NS is to use HIPPARCOS
data on massive stars around the sun as the mass distribution of
progenitors in conjunction with the calculations by \cite{whw02}
for a population synthesis of nearby NS.

The choice of the above representative masses for the BC test is
justified by the results of \cite{Blaschke:2004vq} and \cite{Page:2004fy}
where it was demonstrated that
all objects outside the chosen interval of the masses
$1.1 \div 1.4 M_\odot$ are cooler than those inside this interval.
Even if this systematics is not too strong, this statement
is based on the set of cooling simulations \cite{Blaschke:2004vq} and
is applicable to young objects and can be rephrased as follows:
For a given age the brightest objects are from the most populated mass
interval.

The description of the cooling evolution is mainly given by the equation
combining the energy balance and thermal energy transport \cite{W99}
\begin{eqnarray}
\frac{\partial }{\partial t} (T {\rm e}^\phi)~
&=&-\frac{\epsilon_\nu }{c_V}{\rm e}^{2\phi} + \frac{ {\rm e}^{
\lambda}}{ c_V  r^2} \frac{\partial }{\partial r}\left(\kappa~r^2
{\rm e}^{\phi~+ \lambda}\frac{\partial }{\partial r}\left(T {\rm
e}^{\phi}\right)\right) ~,
\label{Tbal}
\end{eqnarray}
where $\phi$ and $\lambda = - \frac12\ln({1-{2 m}/{r}})$ are
metric coefficients. The heat conductivity $\kappa$, the total
neutrino emissivity $\epsilon_\nu$ and the total specific heat
$c_V/n$ are given as the sum of the corresponding partial
contributions defined for density profiles $n(r)$ of the
constituents of the matter under the conditions of the actual
temperature profile $T(r,t)$. The mass of the star is the
accumulated mass below the surface, $M = m(r=R)$, which together
with the gravitational potential $\phi(r)$ can be determined by
Oppenheimer-Volkov equations (see \cite{Blaschke:2000dy,W99}),
where the energy density profile $\varepsilon=\varepsilon(r)$ and
the pressure profile $p=p(r)$ are defined by the condition of
hydrodynamical equilibrium.
 The boundary condition for the solution of
(\ref{Tbal}) reads $T(r = R_{in},t) = T_{\rm in}(t)$. The energy
flux from the surface has also a contribution from photons
$L_\gamma = 4\pi \sigma R^2 T_s^4$, which is governing the
cooling, when the inner crust temperature $T_{\rm in}$ falls down
to $10^{8}$K.

\begin{figure}
\begin{center}
\psfig{figure=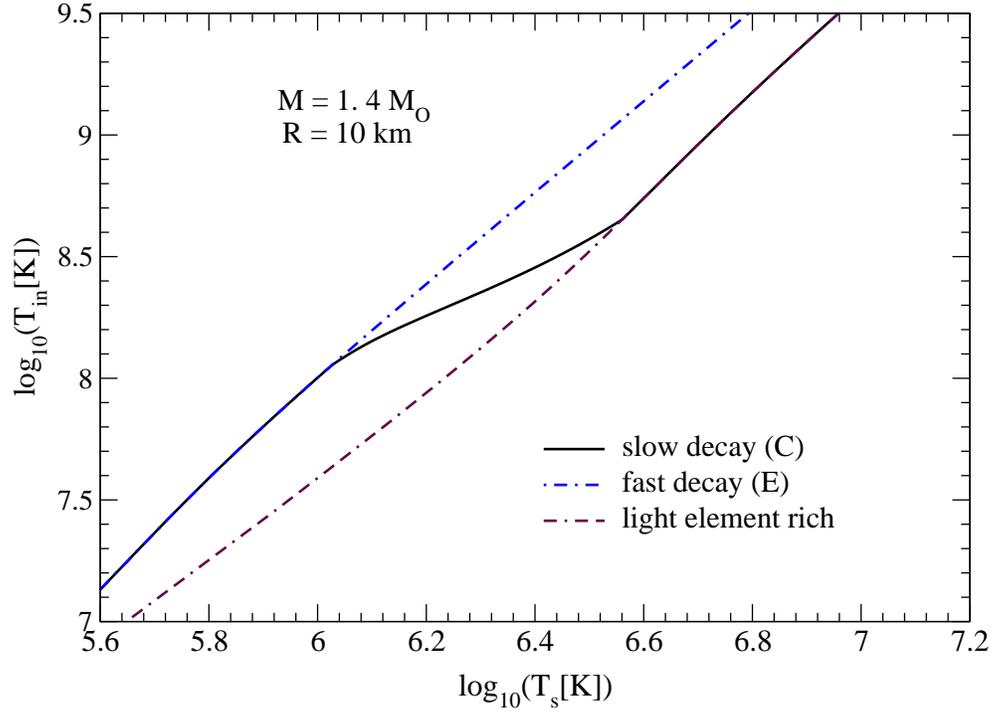,width=0.5\textheight,angle=-90} \caption{
(Color online) The relation between the inner crust temperature
and the surface temperature for different models. Dash-dotted
curves indicate boundaries of the uncertainty band. Notations of
lines are determined in the legend. For more details see
\protect\cite{Blaschke:2004vq,Page:2004fy} and
\protect\cite{Yakovlev:2003ed}. \label{T-in} }
\end{center}
\end{figure}

For different crust models we used in our recent calculations the
$T_{\rm s}-T_{\rm in}$ relations
\cite{Blaschke:2004vq,Grigorian:2005fi,Grigorian:2005ds} shown in
Fig.~\ref{T-in}. The parameter $\eta \sim \Delta M_{\rm L}/M$ is a
measure for the thickness of the light element layer
\cite{Yakovlev:2003ed}, which is related to the pressure at the
bottom of the light element envelope. Thus these borders of the
acceptable $T_{\rm s}-T_{\rm in}$ relations can be denoted as
heavy element $\eta = 4 \cdot 10^{-16}$ (further denoted as crust model (E)
corresponding to the notation in Ref.
\cite{Page:2004fy} ) and light-element $\eta = 4 \cdot 10^{-8}$
crust models, respectively.

  From the discussion we omitted the possible heating in the crust
to make our focus on the general aspect of the choice of crust. In
principle the heating could also be included without
qualitative changes in the argumentation we use here, because it
will change the cooling evolution only in the photon era, but not
the evolution of young objects \cite{VanRiper:1994vp,Tsuruta:2004ue}.

 The surface temperature of NS with thick light element crust
is higher during the neutrino cooling era and it shows slower
cooling for young objects in contrast to the case with a heavy
element rich crust (E) as shown in works \cite{Blaschke:2004vq}
and \cite{Page:2004fy}.

For the cooling simulation done in Refs.
\cite{Blaschke:2004vq,Grigorian:2005fi} we follow the idea of
light element decay, which is in coherence with Ref.
\cite{Page:2004fy}. The assumption is that during the time
evolution of the envelopes chemical composition the mass fraction
of the envelope consisting of light elements decays.
The  time dependence can be described by an exponential
\begin{equation}
\Delta M_{\rm L}(t) = e^{-t/\tau} \Delta M_{\rm L}(0)
\label{decay}
\end{equation}
where $\Delta M_{\rm L}(0)$ is the initial mass of light elements
and $\tau$ is the decay rate.
This decay could be due to the pulsar mechanism which injects
light elements into the magnetosphere or due to nuclear reactions
which convert these elements into heavy ones \cite{CB03a,CB03b}.
This rather complicated picture we had modeled with a fit between
the light and heavy element crust models corresponding to a
{\it slow decay} of light elements (model (C)), whereas the heavy
element crust model is denoted as {\it fast decay} (model (E)),
see Fig. \ref{T-in}.

\begin{figure}[h]
\psfig{figure=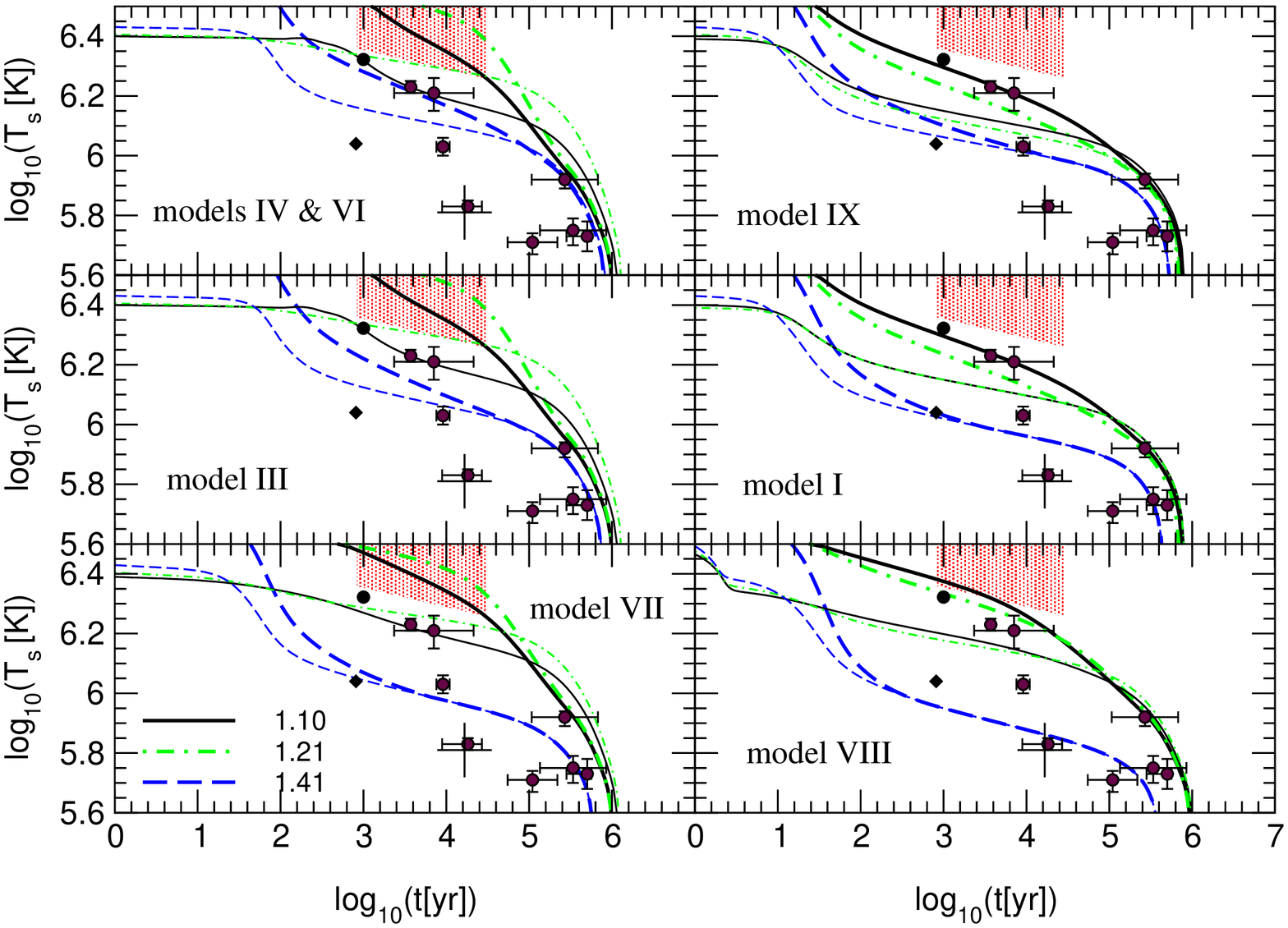,height=0.7\textheight,width=\textwidth,angle=-90}
\caption{(Color online) Cooling evolution for representative NS
from the most populated mass bins $M = 1.1, 1.21, 1.41 M_{\odot}$
according to the NMC scenario. The classification of models is
taken from Ref. \protect\cite{Popov:2004ey}. The cooling curves
for {\it slow decay} (C) are shown with thick lines and those for
{\it fast decay} (E) with thin lines. The data points correspond
to Fig. 1 of Ref. \protect\cite{Page:2004fy}. According to the
{\it brightness constraint} applied for young objects the shaded
regions should not be populated and cooling scenarios entering
there shall be rejected. }
\label{ev}
\end{figure}

For the demonstration of the results given in Fig.~\ref{ev} we
reconsider the same cooling scenarios as in \cite{Popov:2004ey}
using the notation given in Table 1 of that paper.
We exclude only models V and II from our present discussion, since
they are not in our aimed class of models.
The remaining models selected are defined in Table \ref{t1}
by the choice of the nucleon pairing gaps: (A) models I, IX -
gaps from \cite{Takatsuka:2004zq}; (B) models
III, VI, IV, VII, VIII - gaps from \cite{Yakovlev:2003qy}. For all
models the $^3{\mathrm P}_2$ neutron gap is suppressed by
0.1; see Refs.~\cite{Blaschke:2004vq}, \cite{Popov:2004ey} and
\cite{Grigorian:2005fi} for more details.

Each of the six models is calculated with both crust models (C)
and (E). The results are displayed in the six panels of
Fig.~\ref{ev}, grouped into two columns. The left one shows the
models which successfully passed the TA test, but only for crust
model (E) fulfill the BC for younger objects to not have higher
temperatures than those given by the observations. In the right
column all models pass the BC test, while those with crust (E)
failed the TA test leaving some observed points out of
explanation. Note that in Fig. 2 we have not shown cooling curves
for neutron stars with masses exceeding 1.41 $M_\odot$ which
correspond to intermediate or fast coolers since we are focused
here on the discussion of young and slow coolers. We want to
stress, however, that all models in Fig. 2 can pass the TA test if
the crust model is chosen suitably.

  In the upper two panels both models (IV,VI) and IX are calculated
without possible $\pi$-condensation in contrast to the models III
and I, where the condensation is possible. Therefore, the
difference in these corresponding plots is only in the behavior of
the configuration with $M = 1.41~ M_\odot$ for which
$\pi$-condensation occurs. The difference between the four upper
panels (III, IV) and (I, IX) comes from the difference in the
nucleon-nucleon pairing gaps. Only the latter two models are
microscopically justified \cite{Takatsuka:2004zq}.
However, without
additional suppression of the neutron $3P_2$ gap, necessary to reduce the
enhanced cooling by the neutron pair breaking and pair formation
processes, the models would be in disagreement with the TA test (see
\cite{Grigorian:2005fi}). The physical reason of this suppression
could be the medium-modification of the spin-orbit interaction in neutron
matter \cite{Schwenk:2003bc}. Thus the idea of slow decay of light
elements in crust model (C) could be considered as a more
consistent suggestion of the crust-envelope model.

Summarizing our discussion we can conclude the following.
The application of the NMC scenario for the simulation of the cooling
evolution of neutron stars in comparison with the existing observational
data shows that
\begin{itemize}
\item[-] the TA test of cooling scenarios in conjunction with the BC test
can be selective for the discrimination between the crust models,
when nuceon pairing is already chosen, or vice versa;
\item[-]  the TA test of the cooling scenario even with the improvement of the
BC test is not sufficient to make a final
conclusion about the validity of neither the crust model nor the
nuclear superfluidity;
\item[-] since the Log N - Log S test essentially depends on the cooling
behaviour of objects older than $3 \cdot 10^{4}$ yr it does not
interfere with the results of the BC test.
\end{itemize}
All three tests are necessary but not sufficient for the final selection.
With a special choice of the crust model it is possible to change the results
of the TA test while the LogN -LogS test will remain unaffected.
As it is discussed in the Ref. \cite{Gusakov:2004mj} there is another
accelerator of the cooling of low-mass neutron stars, the
direct Urca process of neutrino emission allowed in the mantle of
a neutron star near the crust-core interface, due to inhomogeneous
nuclear structures. Such possibilities will make the crust model of
{\it slow decay} of light elements more preferable.

In conclusion, the present work is a contribution to the development
of general testing schemes for models of compact star cooling
evolution \cite{Popov:2005xa} and strongly interacting matter at
high-densities \cite{Klahn:2006ir} using constraints
from compact star observations.

\section*{Acknowledgement}
I thank David Blaschke for the initiation of this work and his
helpful discussions. Sergei Popov is acknowledged for his comments
to the manuscript and astronomical estimations and Dima
Voskresensky for the critical remarks. The research was supported
in part by the Virtual Institute of the Helmholtz Association
under grant No. VH-VI-041, the DFG Graduiertenkolleg 567 {\it
Stark korrelierte Vielteilchensysteme}, the DFG project No.
436 ARM 17/4/05 and by the DAAD partnership
program between the Universities of Rostock and Yerevan.

\begin{table}
\begin{center}
\begin{tabular}{c||c|c||c}
Class of model  & Models &Models  & Gaps   \\
  & with  $\pi$-condensate  & without $\pi$-condensate & \\
  \hline
A  & I& IX& \cite{Takatsuka:2004zq};~n-$^3{\mathrm P}_2$*0.1  \\
B  & III&  IV \& VI\footnote{The model VI is the same as model IV,
which was already calculated with crust (E) in Ref.~\cite{Blaschke:2004vq}}
 & \cite{Yakovlev:2003qy};~n-$^3{\mathrm P}_2$*0.1   \\
B'  & - &VII& \cite{Yakovlev:2003qy};~p-$^1{\mathrm S}_0$*0.5;~n-$^3{\mathrm P}_2$*0.1   \\
B''  & VIII& - & \cite{Yakovlev:2003qy};~p-$^1{\mathrm S}_0$*0.2;~n-$^1{\mathrm S}_0$*0.5;~n-$^3{\mathrm P}_2$*0.1   \\
\end{tabular}
\end{center}
\caption{Classification of the models by the possible transition
to $\pi$- condensate and the choice of proton
 (p-$^1{\mathrm S}_0$) and neutron (n-$^1{\mathrm
S}_0$,~n-$^3{\mathrm P}_2$) pairing gaps with corresponding
suppression factors.}
 \label{t1}
\end{table}

\begin{thebibliography}{}
\bibitem[\dagger]{address1}
{Permanent address: Department of Physics, Yerevan State
University, 375047 Yerevan, Armenia}
\bibitem{rosat}
 W. Voges et al.,
Astron. Astrophys. {\bf 349} (1999) 389.

\bibitem{Blaschke:2004vq}
  D.~Blaschke, H.~Grigorian and D.~N.~Voskresensky,
  Astron.\ Astrophys.\  {\bf 424} (2004) 979.

\bibitem{Page:2004fy}
  D.~Page, J.~M.~Lattimer, M.~Prakash and A.~W.~Steiner,
  Astrophys. J. Supp. {\bf 155} (2004) 623.

\bibitem{Voskresensky:2001fd}
  D.~N.~Voskresensky,
  Lect.\ Notes Phys.\  {\bf 578} (2001) 467.

\bibitem{SVSWW97}
 Ch.~Schaab, D.~Voskresensky, A.D.~Sedrakian, F.~Weber, and M.~K.~ Weigel,
 Astron. Astrophys.  {\bf 321} (1997) 591.

\bibitem{Blaschke:2000dy}
  D.~Blaschke, H.~Grigorian and D.~N.~Voskresensky,
  Astron.\ Astrophys.\  {\bf 368} (2001) 561.


\bibitem{Schaab:1999as}
  C.~Schaab, A.~Sedrakian, F.~Weber and M.~K.~Weigel,
Astro.\  Astrophys.\ {\bf 346} (1999) 465.

\bibitem{Yakovlev:2004iq}
  D.~G.~Yakovlev and C.~J.~Pethick,
  Ann.\ Rev.\ Astron.\ Astrophys.\  {\bf 42} (2004) 169.

\bibitem{Page:2005fq}
  D.~Page, U.~Geppert and F.~Weber,
  arXiv:astro-ph/0508056 (2005).

\bibitem{Sedrakian:2006mq}
  A.~Sedrakian,
  arXiv:nucl-th/0601086 (2006).

\bibitem{Yakovlev:2003qy}
  D.~G.~Yakovlev, O.~Y.~Gnedin, A.~D.~Kaminker, K.~P.~Levenfish and A.~Y.~Potekhin,
  Adv.\ Space Res.\  {\bf 33} (2004) 523.

\bibitem{p03}
  S.~B.~Popov, M.~Colpi, M.~E.~Prokhorov, A.~Treves and R.~Turolla,
  Astron.\ Astrophys.\  {\bf 406} (2003) 111.

\bibitem{p04}
  S.~B.~Popov, R.~Turolla, M.~E.~Prokhorov, M.~Colpi and A.~Treves,
  Astrophys.\ Space Sci.\  {\bf 299} (2005) 117.



\bibitem{Popov:2004ey}
  S.~Popov, H.~Grigorian, R.~Turolla and D.~Blaschke,
  Astron.\ Astrophys.\  {\bf 448} (2006) 327.

\bibitem{Wilms:2000ez}
  J.~Wilms, A.~Allen and R.~McCray,
  Astrophys.\ J.\ {\bf 542}, (2000) 914.

\bibitem{Popov:2006ma}
  S.~B.~Popov, R.~Turolla and A.~Possenti,
  Mon. Not. Roy. Astron. Soc. {\bf Lett}. (2006) 34.

\bibitem{Faucher-Giguere:2005ny}
  C.~A.~Faucher-Giguere and V.~M.~Kaspi,
  arXiv:astro-ph/0512585.

\bibitem{Kaplan:2004tm}
  D.~L.~Kaplan, D.~A.~Frail, B.~M.~Gaensler, E.~V.~Gotthelf, S.~R.~Kulkarni, P.~O.~Slane and A.~Nechita,
Astrophys.\ J.\ Suppl. {\bf 153} (2004) 269. \\
  D.~L.~Kaplan, B.~M.~Gaensler, S.~R.~Kulkarni and P.~O.~Slane,
  arXiv:astro-ph/0602312.

\bibitem{whw02}
 S.E. Woosley, A. Heger and T. A. Weaver,  Rev. Mod. Phys. {\bf 74}
(2002) 1015.

\bibitem{W99}
F. Weber. {\it Pulsars as Astrophyiscal Laboratories for Nuclear
and Particle Physics},
 ( IOP Publishing, Bristol, 1999).

\bibitem{Grigorian:2005fi}
  H.~Grigorian and D.~N.~Voskresensky,
  Astron.\ Astrophys.\  {\bf 444} (2005) 913.



\bibitem{Grigorian:2005ds}
  H.~Grigorian,
  AIP Conf.\ Proc.\  {\bf 775} (2005) 182.

\bibitem{Yakovlev:2003ed}
  D.~G.~Yakovlev, K.~P.~Levenfish, A.~Y.~Potekhin, O.~Y.~Gnedin and G.~Chabrier,
  Astron.\ Astrophys.\  {\bf 417} (2004) 169.

\bibitem{VanRiper:1994vp}
  K.~Van Riper, B.~Link and R.~Epstein,
  arXiv:astro-ph/9404060 (1994).

\bibitem{Tsuruta:2004ue}
  S.~Tsuruta,
   IAU Symposium 218:{\it  Young Neutron Stars and Their Environment},
  (2004) L18.

\bibitem{CB03a}
P. Chang and L. Bildsten, Astrophys.~J. {\bf 585} (2003) 464.

\bibitem{CB03b}
  P.~Chang and L.~Bildsten,
  Astrophys.\ J.\  {\bf 605} (2004) 830.

\bibitem{Takatsuka:2004zq}
  T.~Takatsuka and R.~Tamagaki,
  Prog.\ Theor.\ Phys.\  {\bf 112} (2004) 37.

\bibitem{Schwenk:2003bc}
  A.~Schwenk and B.~Friman,
  Phys.\ Rev.\ Lett.\  {\bf 92} (2004) 082501.


\bibitem{Gusakov:2004mj}
  M.~E.~Gusakov, D.~G.~Yakovlev, P.~Haensel and O.~Y.~Gnedin,
  Astron.\ Astrophys.\  {\bf 421} (2004) 1143.

\bibitem{Popov:2005xa}
  S.~Popov, H.~Grigorian and D.~Blaschke,
  arXiv:nucl-th/0512098.

\bibitem{Klahn:2006ir}
  T.~Kl\"ahn {\it et al.},
  arXiv:nucl-th/0602038.



\end{thebibliography}
\end{document}